\documentclass[prl,twocolumn, aps,amssymb,footinbib,showpacs]{revtex4-1}

\usepackage{graphicx}
\usepackage{amsmath,amssymb}
\usepackage{cancel}
\usepackage{color}
\usepackage{pifont}
\usepackage{float}

\newcommand{\<}{\langle}

\renewcommand{\>}{\rangle}
\renewcommand{\(}{\left(}
\renewcommand{\)}{\right)}
\renewcommand{\[}{\left[}

\newcommand{\eps}{\epsilon}

\usepackage{ulem}
\usepackage{cancel,ifthen}
\newcommand{\cmnt}[2][NoInPuT]{\ifthenelse{\equal{#1}{NoInPuT}}{}{{\color{red}\sout{#1}}} {\color{blue} #2}}

\definecolor{grey}{rgb}{0.4, 0.4, 0.4}

\usepackage{hyperref}

\begin{document}

\title{Discrete time crystals: rigidity,  criticality, and realizations}
\author{N. Y. Yao$^{1}$, A. C. Potter$^{1,2}$, I.-D. Potirniche$^{1}$, A. Vishwanath$^{1,3}$}
\affiliation{$^{1}$Department of Physics, University of California Berkeley, Berkeley, CA 94720, U.S.A.}
\affiliation{$^{2}$Department of Physics, University of Texas at Austin, Austin, TX 78712, U.S.A.}
\affiliation{$^{3}$Department of Physics, Harvard University, Cambridge MA 02138, U.S.A.}

\begin{abstract}
Despite being forbidden in equilibrium, spontaneous breaking of time translation symmetry can occur in periodically driven, Floquet systems with discrete time-translation symmetry. The  period of the resulting discrete time crystal is quantized to an integer multiple of the drive period, arising from a combination of  collective synchronization and many body localization.
Here, we  consider a simple model for a one dimensional discrete time crystal which explicitly reveals the rigidity of the emergent oscillations as the drive is varied. 
We numerically map out its phase diagram and compute the properties of the dynamical phase transition where the time crystal melts into a trivial Floquet insulator. 
Moreover, we demonstrate that the model can be realized with current experimental technologies and propose a   blueprint based upon a one dimensional chain of trapped ions. 
Using experimental parameters (featuring long-range interactions), we identify the phase boundaries of the ion-time-crystal and propose a measurable signature of the symmetry breaking phase transition. 
\end{abstract}


\maketitle

Spontaneous symmetry breaking---where a quantum  state breaks an underlying symmetry of its parent Hamiltonian---represents a unifying concept in modern physics \cite{peskin1995introduction,chaikin2000principles}. Its ubiquity spans from condensed matter and atomic physics to high energy particle physics; indeed, examples of the phenomenon abound in nature: superconductors, Bose-Einstein condensates, (anti)-ferromagnets, any crystal, and Higgs mass generation for fundamental particles. This diversity seems to suggest that almost any symmetry can be broken.

\begin{figure}
\includegraphics[width=3.in]{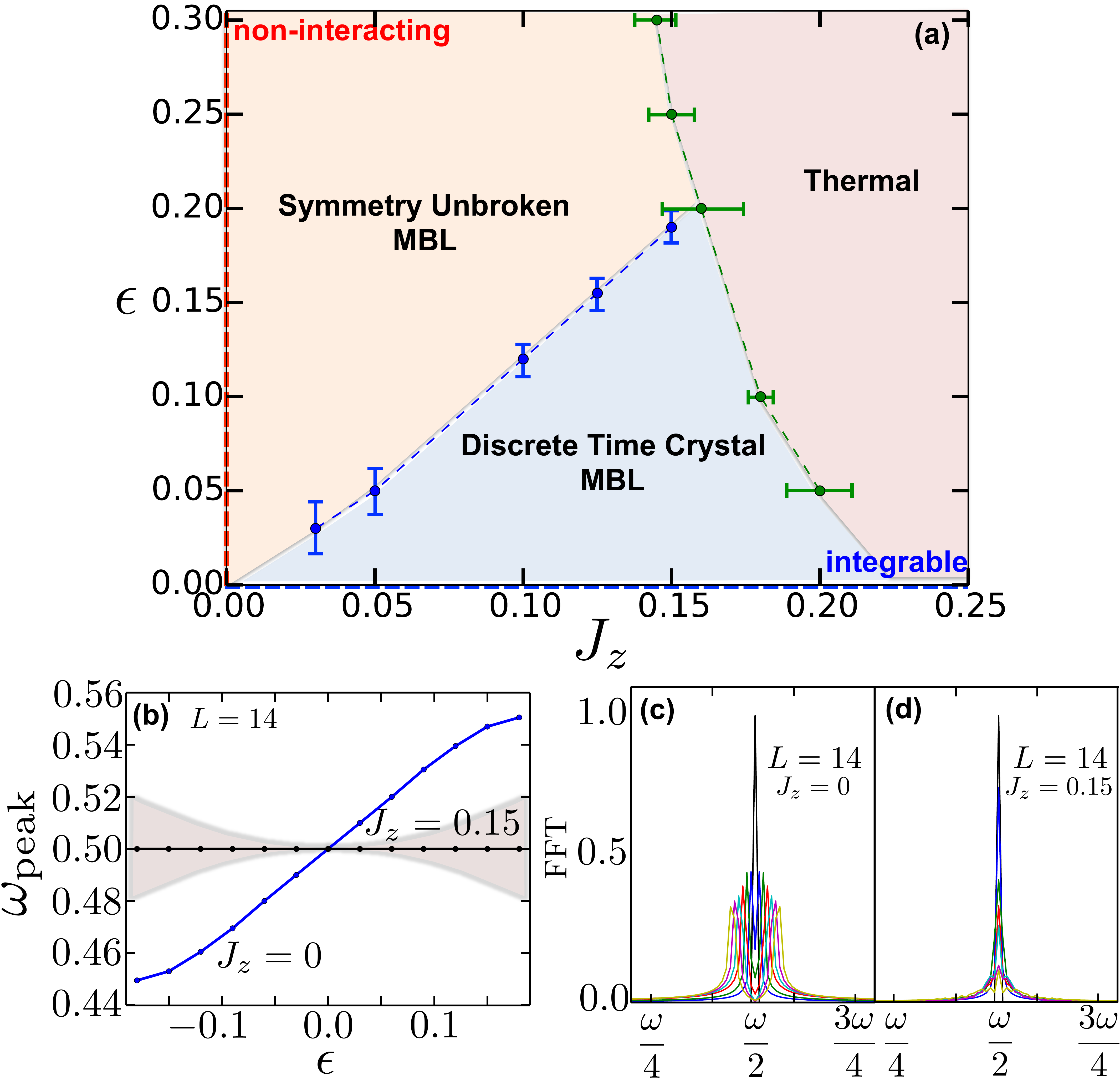}
\caption{%
(a) Phase diagram of the discrete time crystal as a function of interaction strength $J_z$ and pulse imperfections $\epsilon$. (b) Depicts the location of the sub-harmonic Fourier peak as a function of $\epsilon$. In the non-interacting case ($J_z=0$), the peak tracks $\eps$, while in the interacting case ($J_z=0.15$), the peak remains rigidly locked at $\omega/2$. The pink region indicates the FWHM of the  base of the $\omega/2$ peak. Data are obtained at $L=14$ with $10^2$ disorder averages.  (c-d) Representative realizations of the sub-harmonic Fourier response corresponding to $\eps$ in (b). All Fourier transforms are computed using $10<n<150$ Floquet periods.  
}
\label{fig:Intro_new}
\end{figure}

Spurred by this notion, and the analogy to spatial crystals, Wilczek proposed the intriguing concept of a ``time-crystal''---a state which spontaneously breaks \emph{continuous} time translation symmetry \cite{wilczek2012quantum, li2012space, wilczek2013superfluidity}.
Subsequent work developed more precise definitions of such time translation symmetry breaking (TTSB) \cite{bruno2013impossibility,nozieres2013time,volovik2013broken} and ultimately led to a proof of the ``absence of (equilibrium) quantum time crystals'' \cite{watanabe2015absence}. 
However, this proof leaves  the door open to TTSB in an intrinsically out-of-equilibrium setting, and pioneering recent work \cite{khemani2015phase,else2016floquet} has demonstrated that quantum systems subject to periodic driving can indeed exhibit \emph{discrete} TTSB \cite{khemani2015phase,von2016phase,else2016floquet,von2016emergent}; such systems develop persistent macroscopic oscillations at an integer multiple of the driving period, manifesting in a sub-harmonic response for physical observables. 

An important constraint on symmetry breaking in many-body Floquet systems is the need for disorder and localization \cite{khemani2015phase,von2016phase,von2016phase2,else2016floquet,von2016emergent,ponte2015many,ponte2015periodically,lazarides2015fate,abanin2016theory}. In the translation-invariant setting,  Floquet eigenstates are short-range correlated and resemble infinite temperature states which cannot exhibit symmetry breaking \cite{d2014long,lazarides2014equilibrium,ponte2015periodically}.  Under certain conditions, however, prethermal time-crystal-like dynamics can persist for long times \cite{else2016pre,sacha2015modeling} even in the absence of localization before ultimately being destroyed by thermalization \cite{abanin2016theory,heating2}.


In this Letter, we present three main results. First, by exploring the interplay between entanglement, many body localization and TTSB, we produce a phase diagram for a discrete time crystal (DTC) \footnote{We note that this phase is closely related to the $\pi$ spin-glass phase \cite{khemani2015phase,von2016emergent} also referred a Floquet time crystal \cite{else2016floquet}.}.  The DTC, like other symmetry breaking phases, possess macroscopic rigidity and remains locked in its ``collective'' period, displaying a characteristic  `plateau' at the location of its sub-harmonic Fourier response (Fig.~\ref{fig:Intro_new}). This is in stark contrast to free spins, which simply follow the period dictated by the driving. Second, we compute the scaling properties of the dynamical quantum critical point associated with the onset of TTSB, or equivalently the quantum melting of the time-crystal. Third, we propose an experimental realization of the DTC in a one dimensional chain of trapped ions. Using experimental parameters, we identify the phase boundaries of the DTC and propose a measurable signature of the symmetry breaking phase transition.

\emph{Discrete time crystal}---Let us begin by considering a one dimension spin-1/2 chain governed by the binary stroboscopic Floquet Hamiltonian (with period $T = T_1 + T_2$),
\begin{equation}
H_f(t) = \begin{cases}
H_1 \equiv (g - \epsilon) \sum_i \sigma^x_i, &0<t<T_1\\[4pt]
H_2 \equiv\sum_i J^z_i \sigma^z_i \sigma^z_{i+1} + B^z_i \sigma^z_i, \ &T_1<t < T
\end{cases}
\label{eq:model}
\end{equation}
where $\vec{\sigma}$ are Pauli operators, $J^z_i \in [J_z -  \delta J_z, J_z + \delta J_z]$ with $\delta J_z = 0.2 J_z$, and $B^z_i \in [0, W]$ is a random longitudinal field \footnote{It has recently been noted that coupling strength disorder is essential for stabilizing time crystalline order in this model. We thank V. Khemani, C. von Keyserlingk, and S. Sondhi for pointing  this out to us.}.  To simplify the notation, we choose to work in units of~$T_1 = T_2 = 1$, where the Floquet evolution reduces to: $U_f = U_2 U_1 \equiv  e^{-iH_2}e^{-iH_1}$. 
%
%
Throughout the remainder of the paper, we work with $g=\pi/2$ and  note that for generic $\epsilon\neq 0$, the model does not exhibit any microscopic symmetries \cite{von2016emergent}.

\begin{figure}
\includegraphics[width=3.4in]{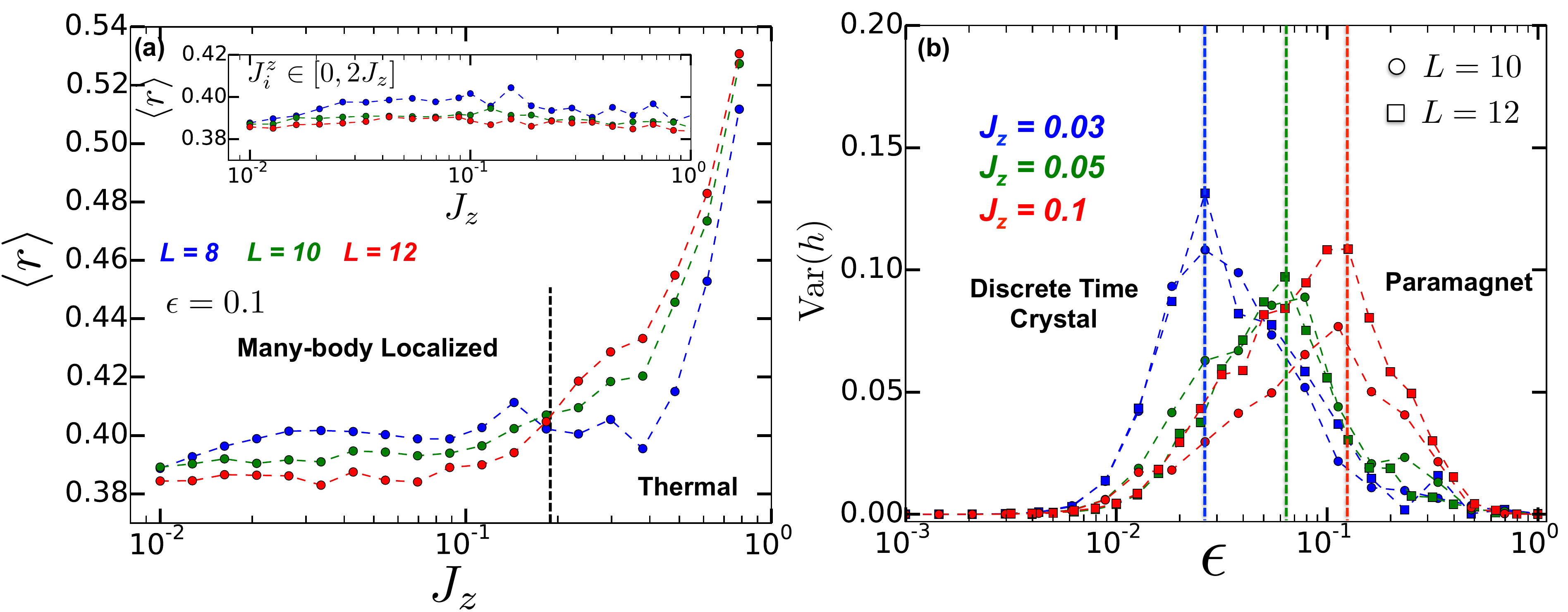}
\caption{%
(a) Level statistics ratio for $\epsilon = 0.1$ as a function of $J_z$. For $L=8$ and $L=10$, we perform $\sim 10^4$ disorder realizations and for $L=12$, we perform $\sim 10^3$ disorder realizations. There is a clear crossing at $J_z \approx 0.18$ indicating the transition. Additional disorder in the interactions, $J^z_i\in[0,2J_z]$, preserves localization over the same parameter range (inset).
(b) Variance of the $\omega/2$ Fourier peak magnitude as a function of $\epsilon$. We observe a clear peak at the transition which exhibits a nearly linear scaling with increasing $J_z$. 
}
\end{figure}

To gain some intuition for the nature of TTSB in this model, let us begin with the ideal decoupled limit where $\epsilon = J_z = 0$. In the parlance of NMR, this simple case corresponds to a chain of decoupled spins undergoing ``spin-echo'' time evolution. To see this, let us consider a random initial product state, $\left| \psi \right \rangle  = \left | \uparrow \downarrow \downarrow \uparrow \downarrow \cdots \right \rangle$, aligned along the $\hat{z}$ direction. The spin-echo unitary, $U_1= e^{-i \pi/2 \sum_i \sigma^x_i}$, flips each spin about the $\hat{x}$-axis, resulting in the oppositely polarized state, $\left| \psi_1 \right \rangle  = \left | \downarrow \uparrow \uparrow \downarrow \uparrow \cdots \right \rangle$. The second unitary results in only a global phase, $\phi$, as each spin is already aligned along $\hat{z}$, $\left| \psi_2 \right \rangle  = e^{i\phi} \left | \downarrow \uparrow \uparrow \downarrow \uparrow \cdots \right \rangle$.  
Since each spin is flipped once per Floquet period, measuring a simple auto-correlation function, $R(t) =\langle \sigma^z_i(t) \sigma^z_i(0) \rangle$, at stroboscopic times (e.g.~$T,2T, \cdots$) yields a perfect train of oscillations \cite{suppinfo}. 
These oscillations imply that $R(t)$ is $2T$ periodic, a fact best captured by its sub-harmonic Fourier response at $\omega/2$---half the binary drive frequency (Fig.~1c). This seems to fit the picture of TTSB and  raises the question: are decoupled spins undergoing ``spin-echo'' a discrete time crystal? The answer lies in the lack of stability to perturbations \cite{winfree1967biological,balents1995temporal,sacha2015modeling}. In this decoupled limit, \emph{any} imperfection in the spin-echo pulse (e.g.~$\epsilon \neq 0$) immediately destroys the $\omega/2$ sub-harmonic. In particular, for $\epsilon > 0$, the unitary, $U_1=  e^{-i (\pi/2 -\epsilon) \sum_i \sigma^x_i}$, leads to beating in $R(t)$ and a splitting of the $\omega/2$ Fourier peak (Fig.~1c). 


\begin{figure}
\includegraphics[width=3.in]{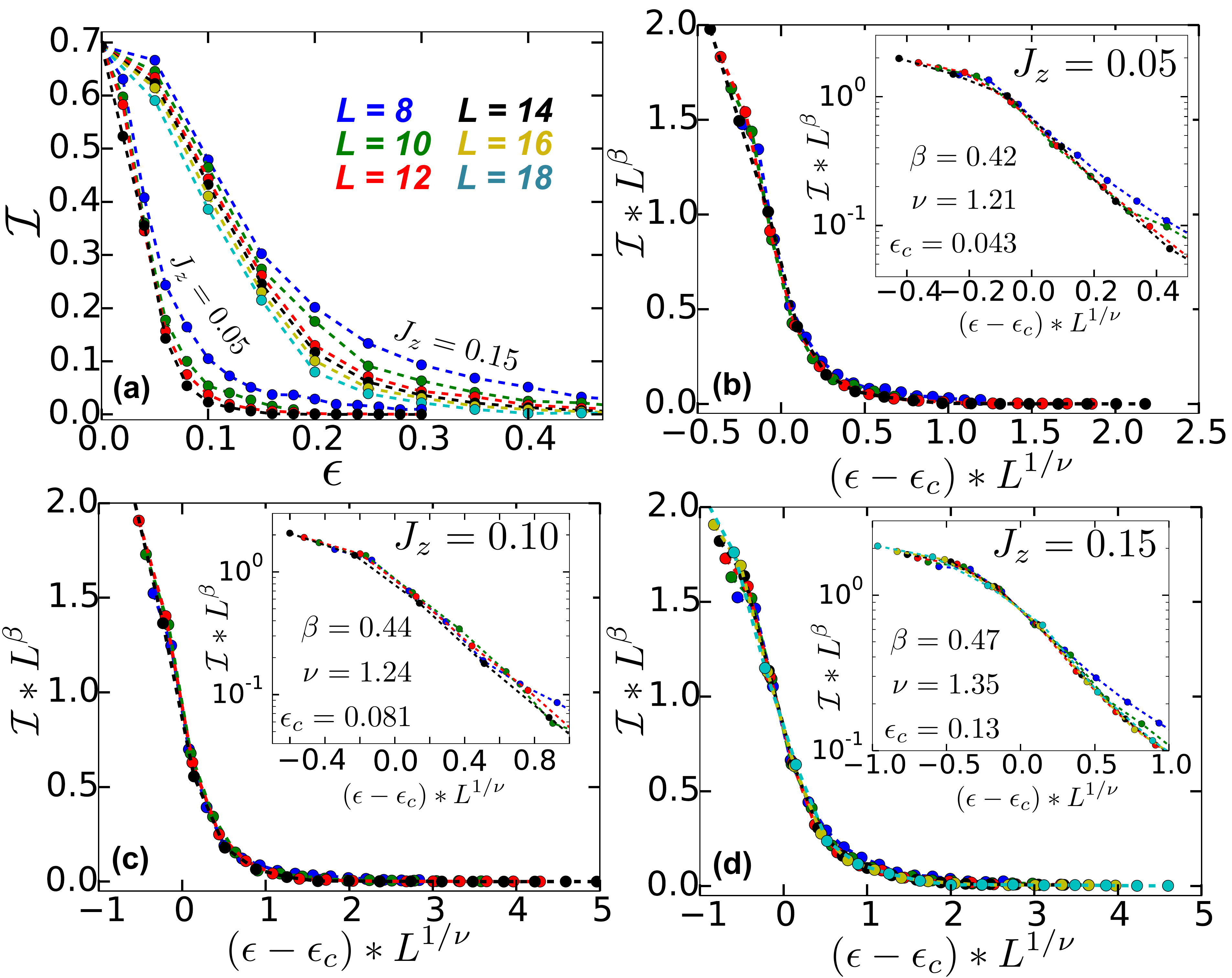}
\caption{(a) Finite size flow of the mutual information between spins on opposite ends of a length $L$ chain. For small detuning, $\epsilon\approx 0$, there is nearly full $\mathcal{I}=\log 2$, long range mutual information, which drops dramatically upon leaving the TTSB phase for large $\epsilon$. (b-d) Scaling collapse of $\mathcal{I}$ to the functional form $\frac{1}{L^{\beta}}f(\frac{L}{|\epsilon-\epsilon_c|^{-\nu}})$
optimized over the parameters $\beta$ and $\nu$. Insets depict the collapsed data with a semi-logarithmic y-axis. Averaging over all $J_z$ yields numerical estimates for the critical exponents: $\beta_{\text{ED}} \approx 0.4\pm 0.1$, and $\nu_\text{ED} \approx 1.3\pm 0.1$; we note that the error bar associated with these exponents ignores the ambiguity in the location of the transition. 
}
\label{fig:scaling}
\end{figure}

Turning on sufficiently strong Ising interaction ($J_z>0$) leads to a qualitatively different story \cite{khemani2015phase,von2016phase,else2016floquet,von2016emergent,else2016pre}. For perfect echo pulses ($\epsilon=0$), the autocorrelation function looks identical to the decoupled case, exhibiting the same normalized Fourier peak at $\omega/2$ (Fig.~1d). Crucially,  imperfections ($\epsilon > 0$) no longer lead to a splitting of the $\omega/2$ Fourier peak, demonstrating the robustness of the system's  sub-harmonic response (Fig.~1d). Herein lies the essence of the discrete time crystal---despite imperfect spin-rotations, collective synchronization from the interactions maintains robust oscillations at half the driving frequency. 
This rigidity is evinced in Fig.~1b \footnote{In Fig. 1b, for $J_z = 0$, $\epsilon \neq 0$, there are two peaks in the disorder averaged Fourier spectra. We compute the splitting, $\delta$, between these peaks and plot $\omega/2 + \delta/2$ at  $+ \epsilon$ and $\omega/2 - \delta/2$ at  $- \epsilon$.}, where the location of the normalized Fourier peak is plotted as a function of $\epsilon$; for finite interactions, this peak is locked at precisely $\omega/2$ 

To explore the phase diagram of the discrete time crystal, we perform extensive numerical simulations to probe both the localization and symmetry breaking phase transitions \cite{disorder}.  We  work at maximal disorder $W = 2\pi$; unlike equilibrium systems, the periodicity of the Floquet unitary limits the strength of the disorder potential.  As the DTC is only  stable in the presence of localization, we begin by characterizing the MBL transition via the quasi-energy level statistics ratio,  $\langle r \rangle = \textrm{min}(\delta_n, \delta_{n+1})/ \textrm{max}(\delta_n, \delta_{n+1})$, where $\delta_n = \mathcal{E}_{n+1} -\mathcal{E}_{n}$ is the $n^{\textrm{th}}$ quasi-energy gap \cite{pal2010many,khemani2015phase}. 
Figure 2a depicts $\langle r \rangle$ as a function of $J_z$ for $\epsilon = 0.1$, where one observes a clear transition at $J_z \approx 0.18$. The evolution of this thermalization transition point for general $J_z$ and $\epsilon$ is shown in Figure 1a (green line). Interestingly, the transition exhibits a weak flow toward larger $J_z$ at small $\epsilon$, consistent with the integrability of $\epsilon = 0$ line.

That the thermalization transition occurs for such weak interactions (more than an order of magnitude smaller than the disorder width) is somewhat surprising; a simple explanation may be that the (imperfect) spin-echo unitary, which  flips each spin by approximately $180^\textrm{o}$, is nearly canceling the random field between the two pieces of the binary drive, leading to effectively weaker disorder. This is consistent with our observation that turning on additional disorder ($J^z_i \in [0,2J_z]$) in the Ising interactions, which are invariant under a uniform spin rotation, leads to a significantly enhanced region of localization (Fig.~2a, inset).


Let us now turn to diagnosing the TTSB transition, which enables us to establish the existence of the discrete time crystal phase and locate its phase boundaries. We will use a combination of four signatures (at infinite temperature): 1) magnitude,  2) variance, and 3) exponential (in system size) persistence---of the sub-harmonic Fourier peak, and 4) mutual information between distant sites \cite{else2016floquet}. We note that a number of other probes of the DTC phase  have also been proposed, including certain eigenstate correlations and responses \cite{von2016emergent}.

 \begin{figure}[t]
 \label{fig:Intro}
\includegraphics[width=2.7in]{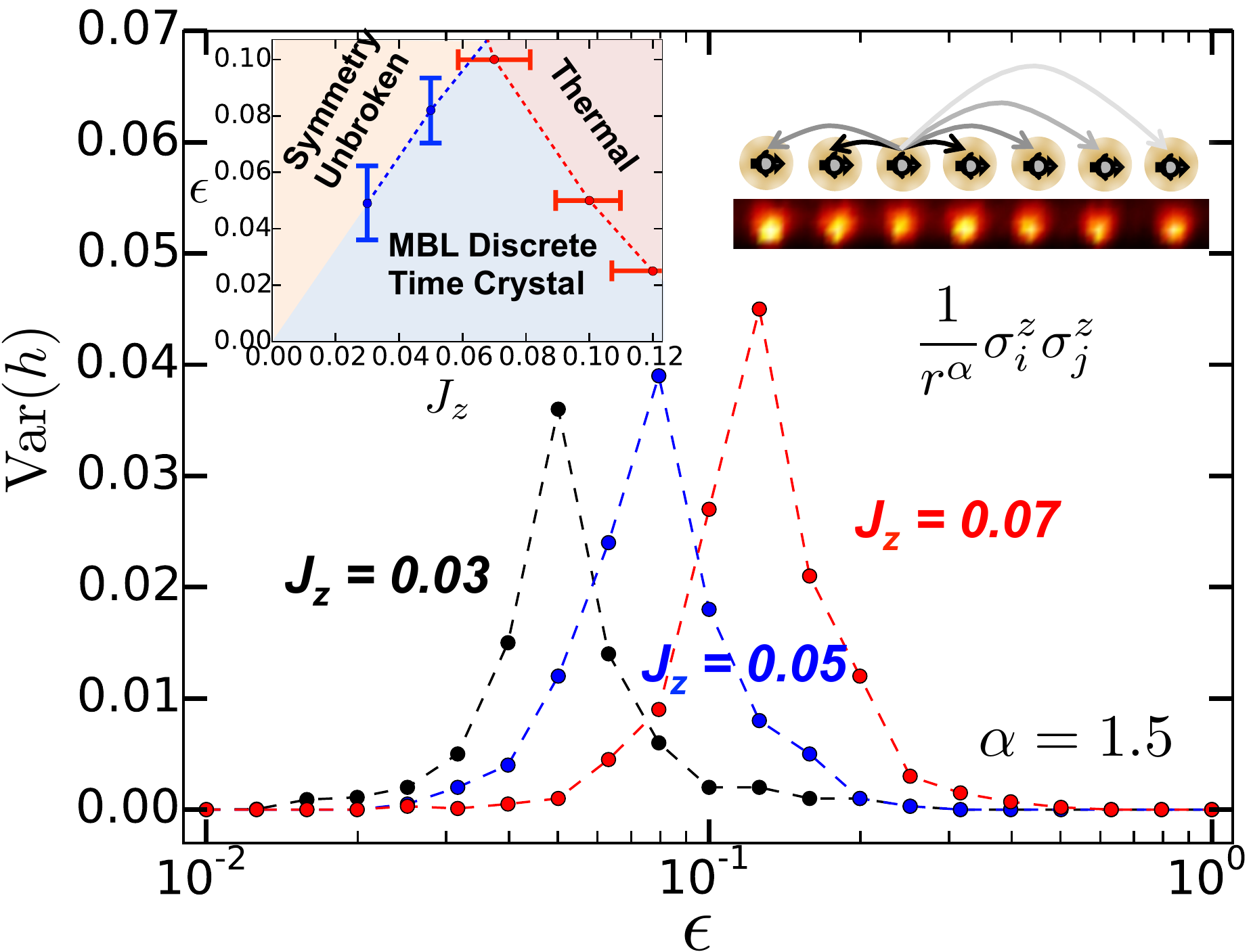}
\caption{Trapped Ion Discrete Time Crystal---Variance of the $\omega/2$ Fourier peak magnitude as a function of $\epsilon$ for power-law Ising interactions with $\alpha = 1.5$ and $L=10$. Unlike the previous case, here, to mimic the experimental scenario, we begin with the same high-energy-density initial state for all simulations, namely, a polarized product state along $\hat{z}$. The location of the TTSB transition can clearly be seen by the peaking of $\text{Var}(h)$. 
The top right inset illustrates a schematic of a one dimensional chain of trapped ions interacting via long-range power law interactions. 
The top left inset depicts the DTC phase diagram for $\alpha = 1.5$. 
}
\end{figure}

We have already encountered the first signature while probing  the rigidity of the $\omega/2$ sub-harmonic response. As one increases the strength of the drive imperfections, $\epsilon$, the magnitude, $h$, of the $\omega/2$ peak decreases (Fig.~1d) and eventually becomes completely washed out when one transitions into the trivial paramagnet \cite{suppinfo}.
The second signature originates from strong critical fluctuations in $h$ near the TTSB transition. This results in a sharp peak in the variance of $h$ and enables one to quantitatively locate the transition in moderate system sizes. As shown in Fig.~2b, increasing $J_z$ strengthens the rigidity of the DTC, shifting the melting transition toward larger detuning, $\epsilon$. We identify the third signature by computing the finite-time-scale where the $\omega/2$ Fourier peak drops below amplitude $0.05$; in the DTC phase, this time scale increases exponentially in system size, while in the trivial phase, it exhibits a significantly weaker dependence \cite{von2016phase,else2016floquet,von2016emergent}. 
The final signature (Fig.~3) relates to the long-range mutual information \cite{else2016floquet} and will be discussed below in the context of the critical scaling properties of the TTSB transition.
As illustrated in Fig.~1a, the combination of these four diagnostics allows us to establish the TTSB transition point as a function of $J_z$ and $\epsilon$ (blue line). 


\emph{Quantum melting transition}---Having mapped out the phase diagram of the DTC, we turn to an analysis of the critical properties of the TTSB transition \cite{suppinfo,husenote}. 
%
%
We obtain the universal scaling properties of this dynamical quantum phase transition by mapping the Floquet evolution to a ``hidden" effective static Ising model whose excited state critical properties can be exactly obtained by renormalization group methods \cite{fisher1992random,vosk2014dynamical,pekker2014hilbert,vasseur2015quantum,you2016entanglement}. Though the TTSB transition falls into the random Ising universality class, we will see that the hidden character of the Ising model introduces notable differences in physical scaling properties. 

For simplicity, our analytic analysis will be performed in  a model where the TTSB transition is tuned via transverse fields instead of spin-echo imperfections (e.g. $\eps=0$)  \cite{von2016phase,else2016floquet,von2016emergent}. While the conclusions will be identical, this approach allows us to compute the effect of $U_1$ exactly and then to treat $U_2$ in a high frequency expansion for $J_zT,|\vec{B}|T\ll 1$. To this end, we consider a modified $H_2 \rightarrow H_2'=\sum_i J_z \sigma^z_{i}\sigma^z_{i+1}+\sum_{\alpha=x,y,z}B^\alpha_i\sigma^\alpha_i$,
where $B^x$ controls the transition and $B^y$ is added to avoid unintentional microscopic symmetries \cite{von2016emergent}.
 
This model exhibits a hidden emergent Ising symmetry $\tilde{S}$ \cite{von2016emergent}, and is in fact, related by a finite depth unitary transformation, $U_\text{FD}$, to a driven transverse field Ising model with $B^{y,z}=0$. In particular, $U_\text{FD}U(T)U_\text{FD}^\dagger =  e^{-iH_\text{TFIM}T}\prod_i\sigma_x^i$, where $H_\text{TFIM}= \sum_i \tilde{J}^z_i \sigma^z_i\sigma^z_{i+1} +\tilde{B}^x_i \sigma_i^x$ has a conventional (e.g. onsite and Hamiltonian independent) symmetry $S = \prod_i \sigma_i^x$, and  $\tilde{J}^z_i$, $\tilde{B}^x_i$ are spatially random quantities, given the disordered character of $U_\text{FD}$. 
%
%

%


To probe the nature of the TTSB transition, our strategy is to consider time-evolution for two Floquet periods,
$
U(2T) = U_\text{FD}^\dagger e^{-2iH_\text{TFIM}T} U_\text{FD}^{\vphantom\dagger}$. Crucially, unlike $U(T)$, this unitary  takes the form of evolution under a \emph{local} transverse field Ising Hamiltonian. Since long-time evolution can always be decomposed into repeated evolutions by $U(2T)$ followed by partial evolution for up to a single period, the late-time properties of the system are governed by those of the excited eigenstates of $H_\text{TFIM}$. For strong disorder, these states exhibit a non-ergodic quantum phase transition between a trivial MBL phase and an Ising symmetry breaking magnetic glass phase. 
%
Thus,  the $Z_2$ discrete time-crystal melting transition, at strong disorder, falls into the universality class of a ``hidden" random Ising transition \cite{fisher1992random,vosk2014dynamical,pekker2014hilbert,vasseur2015quantum}.

A few remarks are in order. The key difference between this ``hidden" Ising transition and the conventional transition is the following: The scaling fields, $\Sigma^\alpha$ of this ``hidden" Ising transition, i.e.~those that exhibit $\overline{\<\Sigma^\alpha(r)\Sigma^\beta(0)\>} = \frac{\delta_{\alpha\beta}}{r^{2\Delta_\alpha}}$ (where the overbar indicates disorder averaging), are related  to those of $H_\text{TFIM}$ by $U_\text{FD}$.  Due to the absence of any microscopic symmetries in the underlying DTC Hamiltonian, the original spins will generically have overlap with \emph{all} scaling fields:
$ \sigma^{\mu}_i = \sum_{i,j,\alpha} c_{ij,\alpha}^{\mu} \Sigma_j^\alpha$ ($\mu=x,y,z$)
where $c_{ij,\alpha}^{\mu} $ are non-universal coefficients that depend on the microscopic details of the lattice and decay exponentially in $|i-j|$. Thus, generic spin-spin correlation functions will also pick up contributions from all scaling fields:
$
\overline{\<\sigma^\mu_i\sigma^\nu_j\> } =  \sum_{i'j'\alpha} c^{(\mu)}_{ii',\alpha}c^{(\nu)}_{jj',\alpha} \frac{1}{|r_{i'j'}|^{2\Delta_{\alpha}}}\approx \frac{1}{|i-j|^{2\Delta_{\alpha_*}}}$.
For large separations, the decay of these correlation functions will be dominated by the scaling field, $\alpha_*$, with the slowest decay (i.e.~minimal scaling dimension, $\Delta_{\alpha_*}$). In the case of the 1D random Ising transition, the magnetization has the slowest decay $\sim 1/r^{\beta}$, where $\beta = 2-\varphi$ and $\varphi = \frac{1+\sqrt{5}}{2}$ is the Golden ratio \cite{fisher1992random}. 

Due to the strong randomness character of the transition, there is a marked difference between the mean scaling behavior just discussed and the typical scaling behavior. 
Indeed, the $1/r^{2-\varphi}$ power-law behavior of \emph{all} local mean  correlation functions results from rare regions that are unusually large, well-ordered and dominate the average \cite{fisher1992random}. Typical correlation functions, on the other hand, all decay significantly faster than any power law, namely, as a stretched exponential: $\<\Sigma^\alpha(r)\Sigma^\beta(0)\>_\text{typ}\sim \delta_{\alpha \beta}e^{-\sqrt{r}}$.
Similarly, the typical and mean scaling properties  will also be governed by two different diverging length-scales: $\xi_\text{typ}\sim |\eps-\eps_c|^{-\nu_\text{typ}}$ and $\xi_\text{avg} \sim |\eps-\eps_c|^{-\nu_\text{avg}}$, with correlation length exponents, $\nu_\text{typ} = 1$ and $\nu_\text{avg} = 2$.

While the above discussion focuses on critical eigenstate properties, in an experiment, one is interested in manifestations of criticality in dynamical signatures.
To this end, one can examine the critical temporal decay of the aforementioned $\omega/2$ Fourier peak. A sharp definition of this mixed time/frequency object can be obtained through the Wigner distribution function: $C^{ab}(\omega_0,t) \equiv \int_0^\infty d\tau e^{-i\omega_0 \tau}\<\sigma^a(t+\tau)\sigma^b(\tau)\>$, which, due to the ``hidden'' Ising structure of the transition will decay asymptotically as the slowest decaying scaling field \cite{vosk2014dynamical}:
\begin{align}
C^{ab}\(\omega_0 = \frac{\omega}{2},t\)\sim \frac{1}{\log^{2-\varphi} t}.
\end{align}
This logarithmically slow decay contrasts with both the power-law decay characteristic of trivial MBL phases, $C_\text{MBL}^{ab} \sim t^{-p}$ and the exponential decay characteristic of a thermalizing system, $C_\text{thermal}^{ab}\sim e^{-t}$ \cite{serbyn2013local}.

\emph{Critical scaling of mutual information}---Having elucidated the scaling structure of the TTSB transition, we now  perform a numerical exploration of the time-crystal-melting transition for the original model [Eqn.~\ref{eq:model}]. In particular,  we compute the mutual information, $\mathcal{I}(L)$, between the first and last site of the spin chain as a function of $\epsilon$ for fixed $J_z$ (Fig.~3) \cite{else2016floquet}. 
%
%
As depicted in Fig.~3a, the mutual information exhibits a clear finite size flow,  sharpening with increasing system size. 
To explore the critical properties of the transition, we  conduct a finite size scaling analysis of this data, based on our analytic understanding of the transition.  In analogy to the disordered Ising transition, the TTSB critical point can be viewed as having a broad distribution of nearly ordered time-crystal clusters. The mutual information between two spins separated by $L$ is of order unity when they belong to the same cluster and exponentially small otherwise. Hence, at criticality, $\mathcal{I}(L)$ tracks the probability for two spins to be in the same cluster, which scales as $\sim L^{-\beta}$ \cite{fisher1992random}. For $\epsilon\approx \epsilon_c$ near the transition, the mutual information will then follow the universal scaling form: $\mathcal{I}\sim \frac{1}{L^{\beta}}f(L/\xi)$, where $\xi\sim |\epsilon-\epsilon_c|^{-\nu}$ is the correlation length of the incipient time-crystal order.

In Fig.~\ref{fig:scaling}b-d, we perform a two parameter scaling collapse on the numerical data for $\mathcal{I}$, by plotting $L^\beta \hspace{0.3mm}\mathcal{I}(L)$ versus $(\epsilon-\epsilon_c)L^{1/\nu}$.
Tuning $\beta$ and $\nu$ to collapse the various system sizes near the critical point, we obtain $\beta_\text{ED} \approx 0.4\pm 0.1$ and $\nu_\text{ED} \approx 1.3\pm 0.1$ (averaged across all interaction strengths $J_z$) \cite{suppinfo}.
These fits are  consistent with the exact analytic expression for $\beta$. The value of $\nu_\text{ED}$ lies  between the expected typical and mean values, likely reflecting the limitations of our small system sizes for capturing rare fluctuations that give $\nu_\text{avg}=2$ in macroscopic systems. 


\emph{Experimental Realization}---We now propose a simple experimental blueprint for the implementation of a discrete time crystal in a one dimensional array of trapped ions \cite{kim2010quantum,blatt2012quantum,smith2015many}. 
In such systems, the spin degree of freedom can be formed from two internal electronic states within each ion; an effective transverse field, $H_T = \Omega \sum_i \sigma^x_i$, can then be realized via resonant microwave radiation between these  electronic states \cite{blatt2012quantum,smith2015many}.
Coulomb repulsion between the ions stabilizes a crystalline configuration and interactions between the spins are generated via off-resonant laser fields that couple each spin with either longitudinal or transverse phonon modes \cite{korenblit2012quantum}.
 This produces long-range Ising-type interactions, $H_{\textrm{int}} = \sum_{ij} J_{ij}/ r_{ij}^\alpha \sigma^z_i \sigma^z_j$, between the spins which fall off as a tunable power-law, with $0<\alpha <3$ (Fig.~4) \cite{korenblit2012quantum,smith2015many}. Finally, a disorder potential can be generated via either individual ion addressing or a 1D optical speckle potential that leads to randomized AC Stark shifts \cite{white2009strongly,kondov2015disorder}.  
In combination, these above ingredients enable the direct realization of a power-law generalization of Eqn.~\ref{eq:model},
\begin{equation}
U^{\textrm{ion}} = \begin{cases}
U_1^{\textrm{ion}} =  e^{-i \Omega \sum_i \sigma^x_i t_1} \\
U_2^{\textrm{ion}} = e^{-i ( \sum_{ij} J_{ij}/ r_{ij}^\alpha \sigma^z_i \sigma^z_j + B^z_i \sigma^z_i )t_2}\\
\end{cases}
\end{equation}
 where the nearest neighbor Ising interaction is replaced by $H_{\textrm{int}}$ and $\{t_1, t_2\}$ represent tunable evolution times. We emphasize that our proposed realization can likewise be naturally implemented in ultracold polar molecules \cite{yan2013observation, hazzard2014many} and Rydberg-dressed neutral atom arrays \cite{zeiher2016many,endres2016cold}, both of which also feature long-range interactions. 

This leads to a key question: can the discrete time crystal  survive the presence of such long-range interactions \cite{burin2006energy,yao2014many,burin2015localization}?
To quantitatively probe the effect of the long-range power law and the existence of a DTC phase in trapped ions, we perform a numerical study of $U^{\textrm{ion}}$ with $\alpha = 1.5$ \cite{longrange}. Diagnosing the MBL transition, one finds that  long-range interactions disfavor localization and the MBL transition shifts significantly toward  smaller $J_z$ (Fig.~1a, red line) \cite{suppinfo}. We note that many-body resonance counting suggests a critical power law, $\alpha_c = 3/2$  in one-dimension \cite{burin2015localization}, although this delocalization is expected to emerge only for very large systems, and we do not find evidence of such critical delocalization in our simulations. 

Interestingly, within the localized phase, power-law interactions seem to better stabilize the DTC phase \cite{suppinfo}.  In particular, starting from a fully polarized product state aligned along $\hat{z}$, we again compute the variance of $h$ as a function of $\epsilon$. As illustrated in Fig.~4, the transition as determined from the peaking of $\textrm{Var}(h)$ is weakly enhanced when compared to the short-range case, leading to a modified phase diagram (Fig.~4, inset).
These results suggest that a trapped ion quantum simulator can naturally realize a discrete time crystal phase, even in the presence of long-range interactions. 
Moreover, within current coherence times \cite{smith2015many}, one can observe $\sim 10^2$ Floquet periods, sufficient to detect both the DTC's sub-harmonic rigidity and to probe its TTSB transition via $\textrm{Var}(h)$.

In summary, we have introduced a simple, one dimensional disordered Floquet system that exhibits a robust discrete time crystal phase. We characterize this phase via several diagnostics including the rigidity of the emergent sub-harmonic frequency to changes/imperfections in the driving. Moreover, we develop a theory of the melting transition from the time crystal into the trivial Floquet paramagnet and utilize this to conjecture a scaling form for the mutual information. Finally, we propose a  realization of the discrete time crystal in a 1D array of long-range-interacting trapped ions and demonstrate that signatures of both the DTC phase and the TTSB transition can be directly observed with current experimental technologies.

We gratefully acknowledge the insights of and  discussions with E. Altman, B. Bauer, P. Hess, D. Huse, V. Khemani, A. Lee, M. Lukin, C. Monroe, C. Nayak, J. Smith, S. Sondhi, C. von Keyserlingk, R. Vasseur, M. Zaletel, J. Zhang.  
We particularly thank V. Khemani, C. von Keyserlingk and S. Sondhi for bringing to our attention an omission in Eqn.~1 in a prior version of the manuscript, which incorrectly left off the presence of coupling strength disorder that was included in the  numerical simulations.
This work was supported, in part by, the AFOSR MURI grant FA9550- 14-1-0035, the Simons Investigator Program,  the Gordon and Betty Moore Foundation's EPiQS Initiative through Grant GBMF4307, and the Miller Institute for Basic Research in Science.

\bibliography{TCBib}

\begin{thebibliography}{55}
\bibitem [{dis()}]{disorder}%
  \BibitemOpen
  \href@noop {} {\bibinfo  {journal} {We work at maximal disorder $W = 2\pi$;
  unlike equilibrium systems, the periodicity of the Floquet unitary limits the
  strength of the disorder potential}\ }\BibitemShut {NoStop}%
\end{thebibliography}

\end{document}


\title{Supplemental Material for Discrete time crystals:  rigidity,  criticality, and realizations}
\author{N. Y. Yao, A. C. Potter, I.-D. Potirniche, A. Vishwanath}
\maketitle

\section{Explicit form of the hidden Ising symmetry \label{app:hiddensymmetry}}
\noindent Here, we explicitly construct the effective Ising symmetry $\tilde{S}$ underlying the discrete time crystal  in a manner that also reveals its physical connection to time-translation symmetry. To diagnose a spontaneous doubling of the time-period, we  compare the difference between the  evolution operator for two periods, $U(2T)$, and that for one period, $U(T)$. In addition to any ``interesting" differences between these operators due to TTSB (e.g. doubling of the spectrum in $U(2T)$), these operators have the trivial difference that any excitation acquires twice as much phase under $U(2T)$ as under $U(T)$. The following procedure removes for this trivial difference. 
First, we note that in the DTC phase, the Floquet evolution for a single period cannot be written as the exponential of a quasi-local Hamiltonian, $U(T)\neq e^{-iHT}$ with $H$ local. However, the evolution for two periods has no such obstacle, and we can define (see Eqn.~(2) maintext): 
\begin{align}
U(2T) = U_\text{FD}^\dagger e^{-2iH_\text{TFIM}T} U_\text{FD}^{\vphantom\dagger}=e^{-2i U_\text{FD}^\dagger H_\text{TFIM} U_\text{FD}^{\vphantom\dagger} T}  = e^{-2iH_\text{eff}T},
\end{align}
with $H_\text{eff}$ being a quasi-local Hamiltonian. This allows us to  use $H_\text{eff}$ to construct the effective symmetry:
\begin{align}
\tilde{S} = U_\text{FD}^\dagger \(\prod_i\sigma_x^i\) U_\text{FD}^{\vphantom\dagger} =  e^{iH_\text{eff}T}U(T),
\end{align}
which precisely captures the difference between $\sqrt{U(2T)}$ and $U(T)$.  
To obtain $e^{iH_\text{eff}T}$, one should take the same branch of the ``square root" for all quasi-energies of $U(2T)$. From this construction, one can immediately identify a number of characteristics of $\tilde{S}$: 1) it commutes with $U(T)$, 2) it squares to the identity, $\tilde{S}^2=1$ and 3) it is a finite depth unitary circuit (since it is constructed from quasi-local Hamiltonian evolutions). Thus, $\tilde{S}$ satisfies all the desired properties of an Ising symmetry.

\vspace{3mm}

\noindent To verify this construction, let us first consider the case of a driven Ising model with $B^{y,z}=0$. Then, one can easily verify that $\tilde{S}=\prod_i\sigma_i^x$ coincides with the ordinary Ising symmetry. For $B^{y,z}\neq 0$, one can explicitly construct $\tilde{S}$ by treating $U_1= e^{-i \pi/2 \sum_i \sigma^x_i}$ exactly and  $U_2' = e^{-i (\sum_i J_z \sigma^z_{i}\sigma^z_{i+1}+\sum_{\alpha=x,y,z}B^\alpha_i\sigma^\alpha_i)} $ in a high frequency expansion, valid in the limit of $J_z,B^\alpha  \ll 2\pi/T$. Through second order in this expansion, $\tilde{S}$ takes the simple form of a randomly rotated Ising symmetry: 
\begin{align}
\tilde{S} &\approx \prod_i e^{-\frac{i}{2}\vec{B}_\perp\cdot\sigma_i}
\sigma_i^x e^{\frac{i}{2}\vec{B}_\perp\cdot\vec{\sigma}_i}+\mathcal{O}\((JT)^3,(|\vec{B}|T)^3\)
\end{align}
where $\vec{B}_\perp$ is the vector containing $B^{y,z}$ but not $B^x$. This simple form of a product of on-site spin rotations  is lost at higher orders in the high-frequency expansion; however, owing to the expected convergence of this expansion in non-thermal MBL phases, higher order corrections should only weakly dress the above expression.

\vspace{3mm}

\noindent {\bf $\Z_n$ Time Crystals}---The above construction can  straightforwardly be extended to more general TTSB phases in which the period is spontaneously $n$-tupled, for arbitrary integer $n>1$. In such cases, we can analogously write the $n$-period Floquet operator, $U(nT) = e^{-inH_\text{eff}T}$, as evolution under a quasi-local Hamiltonian $H_\text{eff}$ and define the $\Z_n$ symmetry operator $\tilde{S} = e^{iH_\text{eff}T}U(T)$. Again,  $\tilde{S}$ will commute with $U(T)$ and satisfies $\tilde{S}^n=1$. This symmetry is broken at the $n$-fold TTSB transition. In systems which have no other symmetries besides this emergent hidden $\Z_n$ symmetry, we may again analyze the transition by looking at the corresponding $\Z_n$ symmetry breaking transition in $H_\text{eff}$. In particular, one can follow the general representation theoretic procedure for constructing RSRG flows for excited states of random 1D chains. Since $\Z_n$ has only one dimensional irreducible representations, the universality class of this $\Z_n$ TTSB transition has the same critical exponents as the random Ising chain for any $n$. Indeed, upon replacing the quantum dimensions and fusion rules  found in \cite{vasseur2015quantum} with the dimensions and fusion rules for the irreducible representations of $\Z_n$, one finds that   the RSRG flow equations become identical to those of the Ising chain, or its dual description in terms of Majorana fermions.

\begin{figure*}[ht]
\begin{center}
\includegraphics[width=0.8\textwidth]{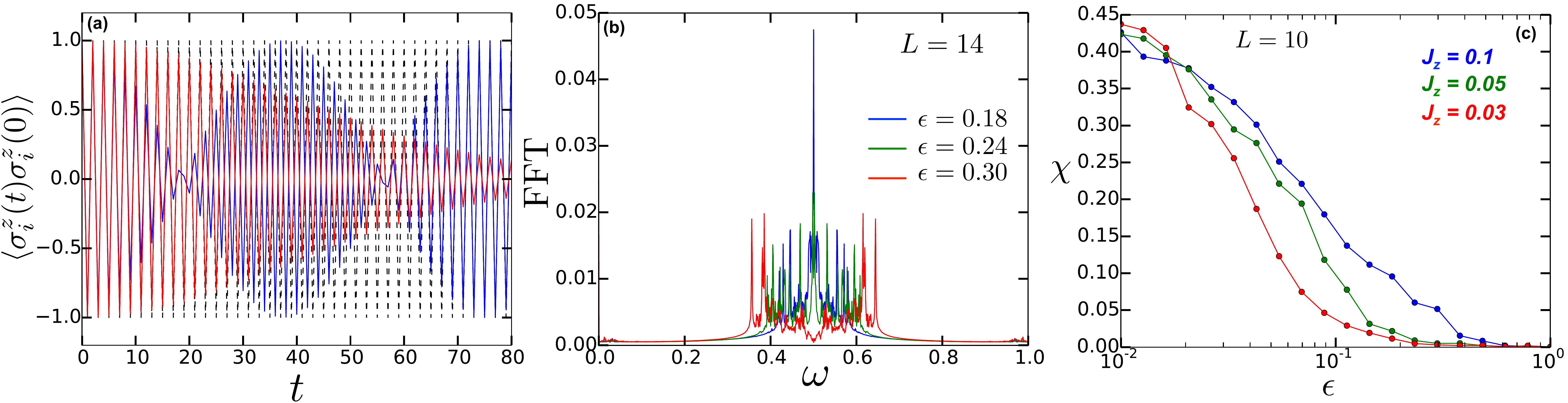}
\end{center}
\caption{ (a) Representative autocorrelation functions, $R(t) = \langle \sigma^z_i(t) \sigma^z_i(0) \rangle_{T=\infty}$  (at stroboscopic times), in a system of $N=14$ spins at maximal disorder:  ideal ``spin-echo''  $\epsilon = 0$ (black dashed line), decoupled, imperfect echo $\epsilon > 0$, $J_z = 0$ (blue line) and  interacting, imperfect echo (red line). 
(b) Depicts the ``washing out'' of the $\omega/2$ Fourier peak as a function of increasing $\epsilon$ for $J_z = 0.15$ at maximal disorder. (c) Shows the susceptibility to TTSB as a function of $\epsilon$.  }
\label{fig:mscheme}
\end{figure*}

\begin{figure*}[ht]
\begin{center}
\includegraphics[width=0.8\textwidth]{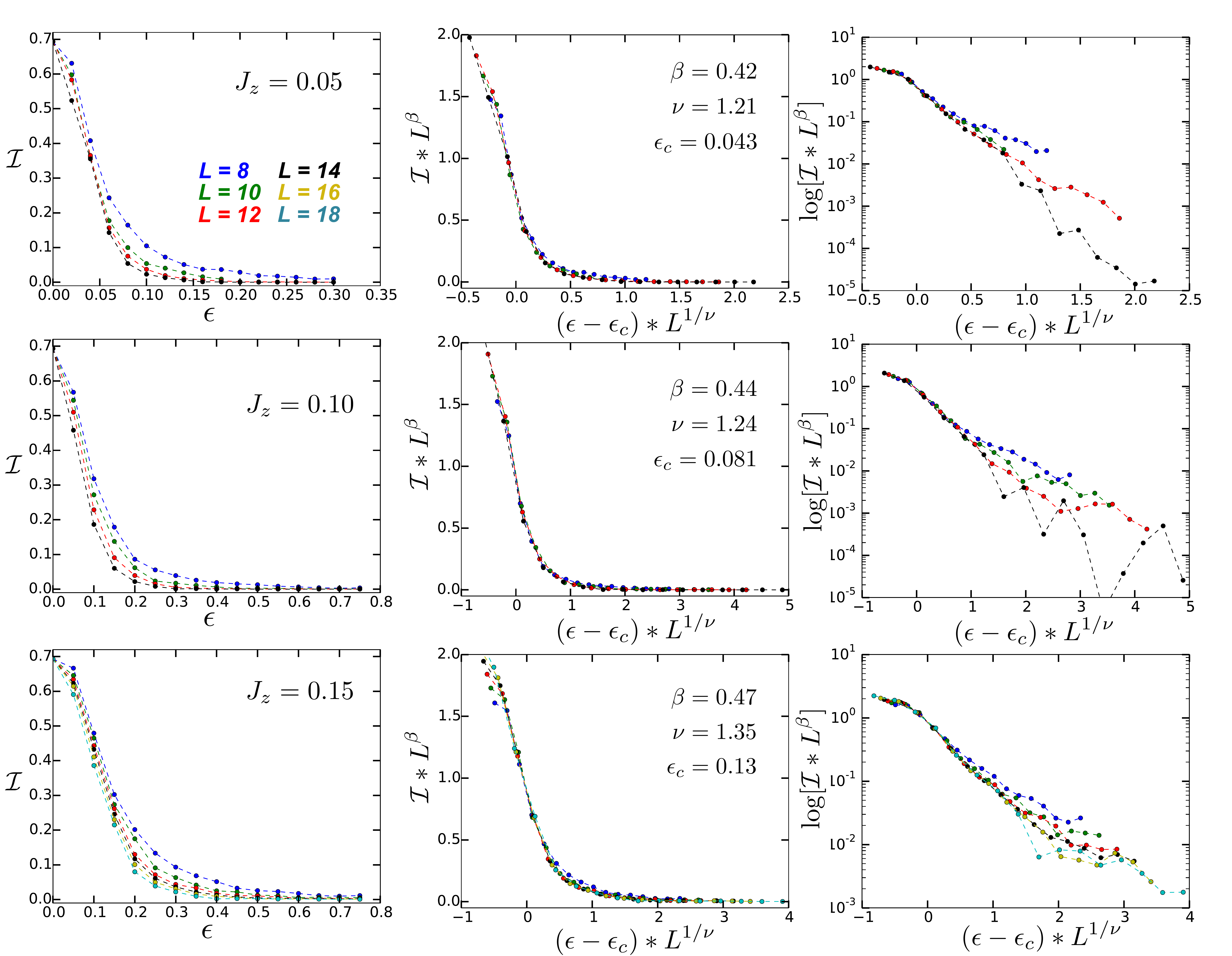}
\end{center}
\caption{ Depicts $\mathcal{I}(L)$ as a function of $\epsilon$ for $J_z = 0.05, 0.1, 0.15$. Across each row, a two parameter scaling is performed on the data  by plotting $L^\beta \hspace{0.3mm}\mathcal{I}(L)$ versus $(\epsilon-\epsilon_c)L^{1/\nu}$ and optimizing the collapse. Extracted optimal parameters are shown. The final column depicts the collapsed data on a semi-logarithmic y-axis.  Near $\epsilon \approx \epsilon_c$, the data exhibits a robust collapse but deviations begin to appear for larger $\epsilon - \epsilon_c$.   }
\label{fig:mscheme}
\end{figure*}

\section{Numerical data and details}

\noindent Here, we provide additional numerical data and simulation details. Figure S1(a) shows a number of representative time traces for the infinite temperature spin-spin auto-correlation function, $R(t) = \langle \sigma^z_i(t) \sigma^z_i(0) \rangle_{T=\infty}$, which captures the various cases described in the maintext. For perfect pulses ($\epsilon = 0$), the correlation function exhibits perfect $2T$ oscillations that continue forever. In the non-interacting case, imperfect pulses lead to a beating of $R(t)$ and hence a splitting of the Fourier peak. In the interacting case, $R(t)$ decays and within the DTC phase, the maximum of the normalized Fourier spectra remains at $\omega/2$. Figure S1(b) illustrates the transition from the DTC to the trivial Floquet insulator as the $\omega/2$ peak vanishes  for large $\epsilon$. Figure S1(c) shows the susceptibility, $\chi$, to time-translation symmetry breaking as a function of $\epsilon$. We compute $\chi$ by considering a doubled Floquet period with the only additions being $\pm \eta \sum_i \sigma^z_i$ to $H_2$. In the first half of the new Floquet evolution, we evolve with $H_2 + \eta \sum_i \sigma^z_i$, while in the second half evolution, we evolve with $H_2 - \eta \sum_i \sigma^z_i$. Evolution via $H_1$ is inter-spliced between as usual.  We work with $\eta = 10^{-4}$ and compute the difference between the eigenstate expectation value of a local $\sigma^z$ correlation function between the original and doubled cases.  We average the absolute value of this quantity across all eigenstates and plot the results versus $\epsilon$ in Figure S1(c). While the trend associated with $\chi$ is consistent with the other diagnostics, the quantitative value of the extracted transition point is approximately a factor of three larger than other diagnostics.
  \begin{figure*}
\begin{center}
\includegraphics[width=1.0\textwidth]{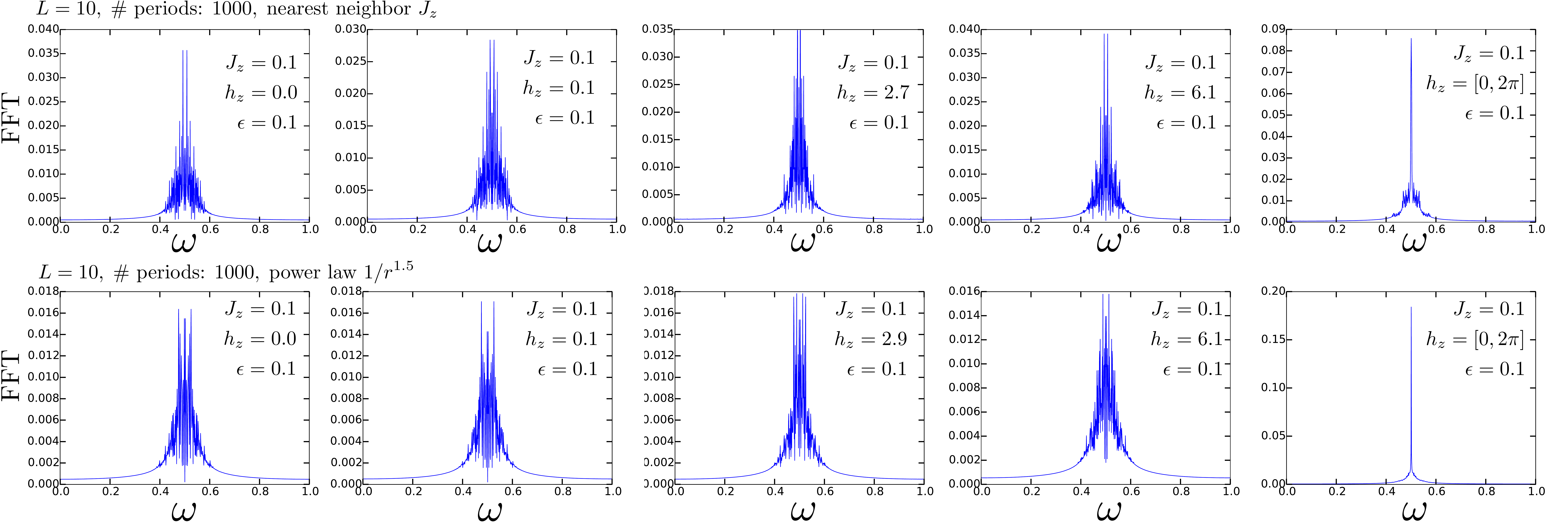}
\end{center}
\caption{ Top row: depicts the sub-harmonic Fourier response of the nearest neighbor model for various uniform fields $h_z$ compared to the disorder-averaged response at maximal disorder in the right most panel.  Bottom row: depicts the sub-harmonic Fourier response of the long-range $\alpha = 1.5$ model for various uniform fields $h_z$ compared to the disorder-averaged response at maximal disorder in the right most panel.  }
\label{fig:mscheme}
\end{figure*}

 Figure S2 shows representative mutual information data at $L=8,10,12,14,16,18$ and $J_z = 0.05,0.1,0.15$.  Figure S3 depicts the differences in the sub-harmonic Fourier response between the clean (undisordered) case and the disordered case. As noted in passing in the maintext, under certain conditions, time-crystal-like dynamics can persist for rather long times even in the absence of localization before ultimately being destroyed by thermalization \cite{abanin2016theory,else2016floquet}. Here, we see that for moderate time-scales corresponding to $n = 1000$ Floquet periods, the clean cases exhibit clear splitting of the sub-harmonic $\omega/2$ peak while the disorder averaged case exhibits a robust single peak. Interestingly, the enhanced power-law stability of the DTC phase (as captured by the increase in the location of the $\text{Var}(h)$ peak in Fig.~4 in the maintext) is also captured by a higher sub-harmonic response and a smaller FWHM.

\bibliography{TCBib}